\documentclass[5p]{elsarticle}
\usepackage{hyperref}
\usepackage{amssymb}
\usepackage{amsmath}
\usepackage[normalem]{ulem}
\usepackage{footnote}
\makesavenoteenv{tabular}
\makesavenoteenv{table}

\usepackage{lineno}

\begin{document}

\begin{frontmatter}
\title{Calibration and performance studies of the balloon-borne hard X-ray polarimeter PoGO+}

\author[KTH,OKC]{M.~Chauvin}
\author[KTH,OKC]{M.~Friis}
\author[KTH]{M.~Jackson\fnref{nowat}}
\author[HU]{T.~Kawano}
\author[KTH,OKC]{M.~Kiss}
\author[KTH,OKC]{V.~Mikhalev\corref{mycorrespondingauthor}}
\author[HU]{N.~Ohashi}
\author[KTH]{T.~Stana}
\author[HU]{H.~Takahashi}
\author[KTH,OKC]{M.~Pearce}

\cortext[mycorrespondingauthor]{Corresponding author.}
\fntext[nowat]{Now at School of Physics and Astronomy, Cardiff University, Cardiff CF24 3AA, UK.}

\address[KTH]{KTH Royal Institute of Technology, Department of Physics, 106 91 Stockholm, Sweden.}
\address[OKC]{The Oskar Klein Centre for Cosmoparticle Physics, AlbaNova University Centre, 106 91 Stockholm, Sweden.}
\address[HU]{Hiroshima University, Department of Physical Science, Hiroshima 739-8526, Japan.}

\begin{abstract}
Polarimetric observations of celestial sources in the hard X-ray band stand to provide new information on emission mechanisms and source geometries. PoGO+ is a Compton scattering polarimeter (20--150 keV) optimised for the observation of the Crab (pulsar and wind nebula) and Cygnus X-1 (black hole binary),  from a stratospheric balloon-borne platform launched from the Esrange Space Centre in summer 2016. Prior to flight, the response of the polarimeter has been studied with polarised and unpolarised X-rays allowing a Geant4-based simulation model to be validated. The expected modulation factor for Crab observations is found to be $M_{\mathrm{Crab}}=(41.75\pm0.85)\%$, resulting in an expected Minimum Detectable Polarisation (MDP) of $7.3\%$ for a 7 day flight. This will allow a measurement of the Crab polarisation parameters with at least $5\sigma$ statistical significance assuming a polarisation fraction $\sim20\%$ -- a significant improvement over the PoGOLite Pathfinder mission which flew in 2013 and from which the PoGO+ design is developed.
\end{abstract}
\end{frontmatter}
\section{Introduction}

Compact celestial X-ray sources such as neutron stars, black hole binaries and active galactic nuclei are routinely characterised by studying the spatial, spectral and temporal properties of their high-energy emissions. Progress in the field is currently driven by such observations. 
Since the transfer of non-thermal radiation in asymmetric systems is intrinsic to many compact sources, the high-energy emission may be polarised.
The addition of two observables --- the linear polarisation fraction ($\Pi$) and corresponding polarisation angle ($\psi$) can add significant diagnostic value, allowing geometrical and physical effects to be disentangled for sources which cannot be spatially resolved, e.g.~\cite{motivation1,motivation2}.
While polarimetry has long been a probe of sources within radio, optical and infra-red astronomy, the application to high energy astronomy has not evolved as rapidly, in part due to the positive definite nature of polarimetric measurements which requires the use of dedicated instrumentation for which the polarimetric response is studied prior to launch. 
  
A measurement of the polarisation of X-rays from the Crab nebula was reported over 40 years ago~\cite{OSO8}. There has been little progress since then, although polarimetric measurements of the Crab and Cygnus X-1 have been recently presented using the IBIS and SPI instruments (not designed as dedicated polarimeters, nor calibrated as such) on-board the INTEGRAL satellite~\cite{Forot,Chauvin,Dean,Laurent,Jourdain,Moran}. Until dedicated satellite-based polarimeters are realised\footnote{e.g. the XIPE mission~\cite{XIPE} (4--8~keV energy range) which is currently competing for selection as the fourth European Space Agency Medium-class mission.}, it is possible to make progress in the field using instruments flown on-board stratospheric balloon-borne platforms at altitudes of $\sim$40~km, representing an atmospheric overburden of $\sim$5~g/cm$^2$. Following a flight conducted in 2013, a balloon-borne mission, the PoGOLite Pathfinder~\cite{ppdescription}, reported a measurement of the polarisation of Crab emissions in the energy range, 20--120~keV~\cite{ppresults}. Due to technical difficulties during the flight (primarily due to inadequate thermal modelling of the payload) only modest statistical precision was attained, $\Pi=(18.4^{+9.8}_{-10.6})\%$; $\psi=(149.2\pm16.0)^\circ$.

The design of the PoGO+ balloon-borne polarimeter is based on the experience gained during the PoGOLite Pathfinder mission, resulting in a polarimeter with significantly improved performance, allowing observations of bright sources such as the Crab and Cygnus X-1 at high statistical significance. 
With sufficient statistics, contributions from the Crab pulsar and nebula may be separated through phase selections on the light-curve.  
For the synchrotron processes at play in the Crab pulsar and wind nebula, the electric field vector of the X-ray flux is perpendicular to the magnetic field lines in the emitting region and hence a polarisation measurement determines the direction of magnetic field and can identify the emission site. In the hard spectral state of Cygnus X-1 (low accretion rate), the inner part of the accretion disk comprises an optically thin hot flow. Thermal X-rays up-scattered from electrons in the hot flow may Compton scatter from the cooler accretion disk yielding polarised emission. Since the electric field vector is perpendicular to the plane of scattering, a polarisation measurement determines the geometrical relation between the photon source and the scatterer. 

PoGO+ was launched from the Esrange Space Centre in northern Sweden on July 12th, 2016. The flight was terminated over Victoria Island, Canada, after 7 days. The flight proceeded as foreseen, resulting in daily observations of the Crab and Cygnus X-1. The purpose of this paper is to detail the polarimetric performance of PoGO+ derived from Monte Carlo simulation studies in the Geant4 framework~\cite{GEANT4} and describe calibration tests with radioactive sources conducted prior to flight. The design of the polarimeter is presented in the next section and is followed by a discussion of the data acquisition system. In Section~\ref{sec:sim} validation studies of the simulation model are described. Section~\ref{sec:response} presents laboratory measurements of the response of the polarimeter to polarised and unpolarised radiation, including expected polarimetric performance for the Crab and Cygnus X-1.

\section{Polarimeter design}

PoGO+ is a Compton scattering polarimeter, where the azimuthal scattering angle is measured for a constrained range of polar scattering angles. For Compton scattering, 
the Klein Nishina differential cross-section dictates that 
X-rays are more likely to scatter in the direction perpendicular to the polarisation vector. The resulting harmonic distribution, $f(\phi)$, of azimuthal scattering angles is referred to as a modulation curve,
\begin{equation}
f(\phi) = N (1 + A \cos(2(\phi - \phi_0)))
\label{eqn:modcurve}
\end{equation}
where A is the modulation amplitude, N is the number of events and $\phi_0$ is the phase of the modulation, which relates to the polarisation angle, $\psi = \phi_0 - 90^\circ$. 
Assuming that background contributions are not polarised, the modulation factor is defined as, 
\begin{equation}
M =A(1+1/R) 
\end{equation}
where R is the signal-to-background ratio for observations. For the PoGOLite Pathfinder flight, $R=0.25$. Monte Carlo simulations show that the value for PoGO+ will be comparable with a dominant background contribution due to atmospheric neutrons. A significant increase in the value of $A$ is sought for PoGO+.

The reconstructed polarisation fraction is defined as $p = M/M_{100}$, where $M_{100}$ is the modulation factor for a 100\% polarised incident beam with defined spectral properties, derived from Geant4 computer simulations benchmarked against laboratory measurements. 
The polarisation parameters can be extracted by fitting the modulation curve. Stokes parameters may also be used which avoids binning subjectivity. 
An established figure-of-merit for X-ray polarimeters is the Minimum Detectable Polarisation (MDP)~\cite{MDP}, defined at 99\% confidence level as
\begin{equation}
\mathrm{MDP} = \frac{4.29}{M_{100} R_s} \sqrt{\frac{R_s + R_b}{T}}
\end{equation}
where $R_s$ ($R_b$) is the signal (background) rate in Hz for polarisation events and $T$ is the observation time in seconds.
For an unpolarised flux there is a 1\% probability of measuring a polarisation fraction greater than the MDP due to a statistical fluctuation.
The aim of PoGO+ is to achieve $\mathrm{MDP}\lesssim10\%$ for Crab observations during a single balloon flight.

\begin{figure}[!th]
 \centering
    \includegraphics[width=0.5\textwidth]{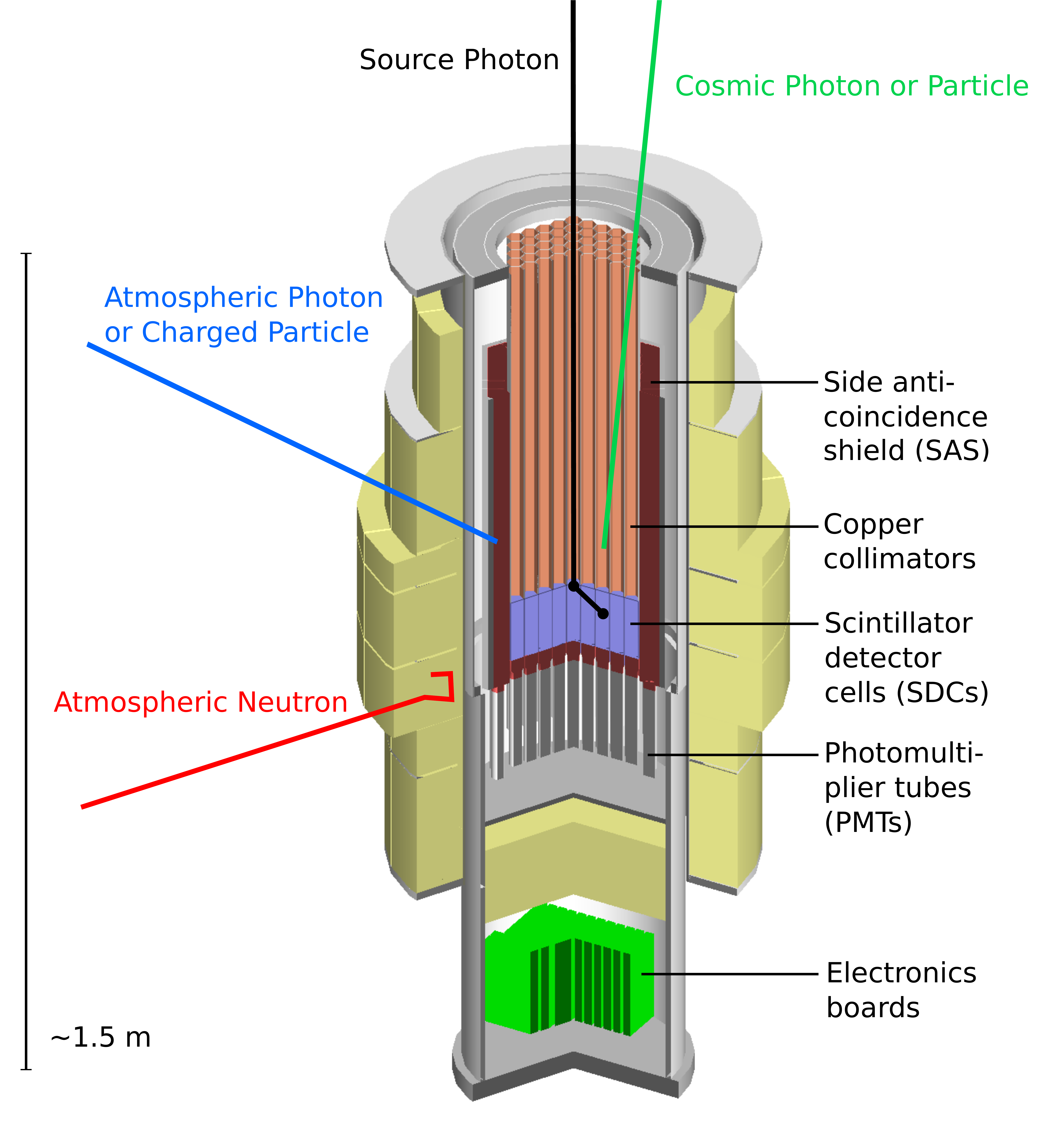}
\caption{A schematic overview of the PoGO+ polarimeter. The detector array comprises 61 scintillator detector cells (SDCs). Each plastic scintillator (blue), is coupled to a BGO element (red) and a copper/tin/lead collimator (orange). The plastic scintillator is EJ-204 from Eljen Technology and the BGO element is provided by the Nikolaev Institute of Inorganic Chemistry. The BGO element connects to a modified Hamamatsu Photonics R7899 photomultiplier tube (grey). The array of SDCs is placed in a segmented lateral BGO anti-coincidence system (SAS) shown in red. Data acquisition electronics are shown in green. The polarimeter components are mounted in a cylindrical pressure vessel which rotates around the viewing axis. The pressure vessel is surrounded by a stationary shield of polyethylene (yellow) to reduce the neutron background. Polyethylene shielding also fills the volume between the electronics and the PMTs.}
  \label{fig:design}
\end{figure}

A schematic overview of the PoGO+ polarimeter is shown in Figure~\ref{fig:design} and its main characteristics are summarised in Tables~\ref{tab:overall} and~\ref{tab:scintillators}.
Azimuthal scattering angles are measured in a close-packed array of 61 hexagonal cross-section plastic scintillator rods (EJ-204, 12~cm long, side length of 2.8~cm) with a scintillation decay time of 1.8~ns. Polarisation events are characterised by a Compton scatter in one rod followed by a second Compton scatter or a photoelectric absorption in another rod within a coincidence window of 110~ns. Aperture background is reduced by a collimator system comprising hexagonal cross-section copper tubes (67.5~cm long) which are mounted immediately in front of each scintillator rod. The copper tube (0.5~mm wall thickness) is covered in foils of lead (100~$\mu$m) and tin (100~$\mu$m) in order to mitigate fluorescence X-rays. Monte Carlo simulations show that the field-of-view of the instrument is $\sim 2^\circ \times 2^\circ$, defined by a 50\% reduction in source flux. 

\begin{table*}[t]
\begin{center}
\begin{tabular}{ |c|c|c|c|c|c|c| }
\hline
Energy (keV) & 20 & 35 & 60 & 100 & 150 & Total/Average\\
\hline
\hline
Collimator opening angle & & & & & & $\sim 2^\circ \times 2^\circ$ \\
Geometric area (cm$^2$) &  & & & & & 378\\
Effective area (cm$^2$) & 0.17 & 6.9 & 7.2 & 10.2 & 9.5 & \\
Modulation factor for Crab (\%)  & 36.3 & 38.0 & 44.1 & 54.8 & 50.7 & 41.75\\
\hline
\end{tabular}
\caption {Overall properties of the polarimeter. The effective area includes flux attenuation by the atmosphere.}
\label{tab:overall}
\end{center}
\end{table*}

\begin{table*}[t]
\begin{center}
\begin{tabular}{ |c| c c c| }
\hline
Scintillator & Plastic scintillator & Bottom BGO & SAS \\
\hline
\hline
Type & EJ-204\footnote{Provided by ELJEN Technology, Sweetwater, Texas, USA.} & \multicolumn{2}{c|}{BGO\footnote{Provided by the Nikolaev Institute of Inorganic Chemistry, Novosibirsk, Russia.}} \\
Decay time (ns) & 1.8 & \multicolumn{2}{c|}{300} \\
Density (g cm$^{-3}$) & 1.023 & \multicolumn{2}{c|}{7.13} \\
Peak wavelength emission (nm) & 408 & \multicolumn{2}{c|}{480} \\
Height (cm) & 12 & 4 & 60 \\
Area (cm$^2$) & 6.2 & 6.2 & 10\\
Sensitivity (photo-electrons keV$^{-1}$) & $0.87\pm0.08$ & $0.38\pm0.03$ & $0.11\pm0.01$ \\
\hline
\end{tabular}
\caption {Scintillator properties. The sensitivity is calculated for an assembled SDC (or SAS unit) thus including the PMT response and light attenuation.}
\label{tab:scintillators}
\end{center}
\end{table*}

A 4 cm tall BGO crystal is glued to the base of each plastic scintillator, providing bottom anticoincidence as well as an interface to a photomultiplier tube (PMT). This assembly is refered to as a scintillator detector cell (SDC). Pulse shape discrimination is applied to the digitised PMT signals in order to distinguish contributions from the plastic and BGO scintillators. To address out-of-acceptance background, the SDC array is housed within a side anticoincidence shield (SAS), which comprises 30 BGO assemblies of length 60 cm and thickness between 3 cm (sides) and 4 cm (corners). Two independent LiCAF-based neutron monitors~\cite{PogolinoNeutron} are mounted in the vicinity of the scintillator array for background monitoring during flight. 

The detector volume is rolled back-and-forth (1$^\circ$ s$^{-1}$ over 360$^\circ$) around the viewing axis in order to mitigate instrument systematics such as variations in efficiency between SDCs. It is surrounded by a stationary polyethylene neutron shield (15 cm thick at the location of the scintillator target). 
An attitude control system~\cite{ppdescription} is used to centre the field-of-view of the polarimeter on observation targets during flight.   
The polarimeter (630 kg), attitude control system (300 kg) and ancillary equipment such as power supplies and communications equipment are housed in a protective gondola assembly (800 kg) provided by the Esrange Space Centre. The flight configuration including the ballast weighs 2200 kg.

The PoGO+ design was developed through computer simulations~\cite{pogo+} based on experience from the PoGOLite Pathfinder flight. 
The primary differences between the PoGO+ and Pathfinder polarimeter designs are:
\begin{enumerate}
\item For the PoGOLite Pathfinder, the collimator system comprised thin-walled (2~mm) hexagonal cross-section tubes of plastic scintillator with a decay time of 285~ns. These active collimators were 60~cm long and interactions occurring far from the PMT could not be reliably distinguished from low energy Compton scattering events in the plastic scintillators, which have a decay time of 1.8 ns (a 25~keV X-ray scattering through a polar angle of 90$^\circ$ will deposit $\sim$1.2 keV).  The use of passive collimators increases the efficiency of pulse shape discrimination between the plastic scintillator and BGO. Furthermore, the detector sensitivity is increased from ($0.44\pm0.05$) to ($0.87\pm0.08$) photo-electrons keV$^{-1}$  as the plastic scintillator can be more completely wrapped in reflective materials. This decreases the low energy limit for measurements. The effective area is also increased since the passive collimators have thinner walls. There is also an increase in the live-time since there is no longer a length of scintillator collimator extending past the side anti-coincidence shield.

\item The length of the scattering scintillator is reduced from 20~cm to 12~cm, thereby improving light collection efficiency, as well as  reducing the background rate. Due to the inverse power-law nature of source spectra ($dN/dE \sim E^{-2}$), the majority of photons interact in the top part of the scintillator and the signal rate is not affected. The shorter scintillator also constrains the range of  
polar scattering angles, yielding a higher $M_{100}$.  A lower total rate leads to an increased live-time. The combination of these factors significantly improves the MDP.

\item Coatings with a higher UV reflectivity were used on all scintillators. Most importantly, Vikuiti Enhanced Specular Reflector film (ESR) was used rather than VM2000 for the plastic scintillator\footnote{Vikuiti ESR and VM2000 are specular reflective foils manufactured by 3M. VM2000 is now obsolete.}.  Prior to integration in the polarimeter, all scintillators were thoroughly tested for optical isolation in order to avoid the cross-talk observed between detector cells in the Pathfinder polarimeter.

\item The hermeticity of the polyethylene neutron shield was improved by overlapping the bottom of the lateral shielding with the shielding located below the PMTs. 

\end{enumerate}

\section{Data acquisition system}
\label{daq}

Balloon flights from Esrange occur at relatively high magnetic latitude, resulting in high background rates (several hundred kHz) compared to the expected source signal rate (around 2 Hz). Since an event read-out induces approximately 10 $\mu$s dead-time, the majority of events cannot be recorded and a sophisticated veto system is therefore needed, which dictates the design of the data acquisition system (DAQ).

\subsection{Electronics}

\begin{figure*}[!th]
  \centerline{\includegraphics[width=1.0\textwidth]{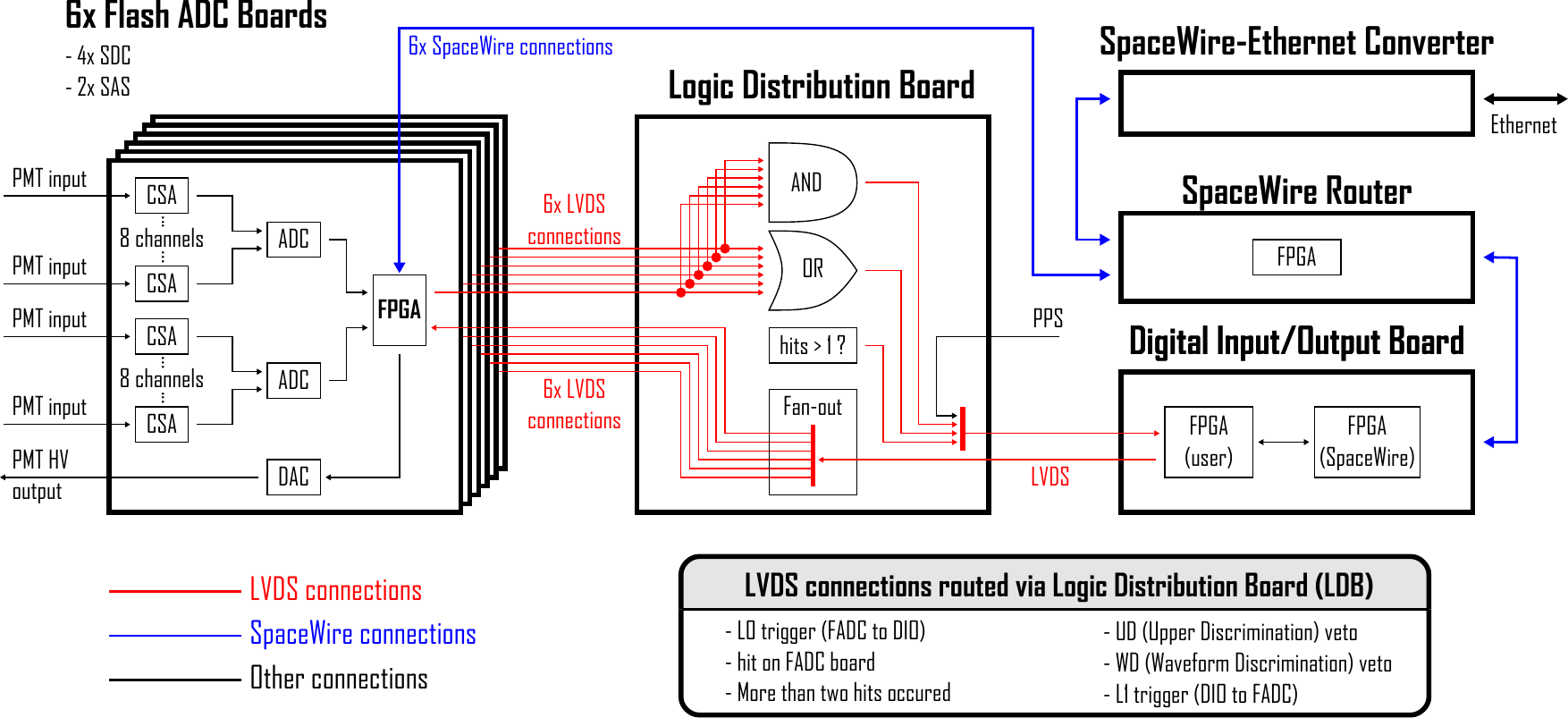}}
  \caption{Block diagram of the DAQ system. PMT signals from the scintillator detector cells (SDCs) and the side anti-coincidence shield (SAS) are fed to FADC boards where candidate events are identified in the FPGA of the FADC and may issue an L0 trigger or veto signals which are routed to a Logic Distribution Board (LDB) and on to a Digital Input Output (DIO) board. Additionally, the LDB allows the overall hit multiplicity of the event to be determined by either forwarding a multiple-hit event received from one FADC or counting the number of single-hit events from multiple FADCs. In the absence of veto signals the DIO issues an L1 trigger.  Each FADC board houses digital-to-analog converters (DACs) which control the high-voltage bias of each connected PMT.  The DAQ system is configured using Ethernet protocols which are converted to SpaceWire~\cite{SpaceWire} - the communications protocol adopted between the FADC and DIO boards, which are interconnected through a router board. The LDB distributes a PPS timing signal from the GPS system in the attitude control system to the DIO.}
   \label{fig:daq-diagram}
\end{figure*}

The DAQ is housed within a cooled enclosure located at the bottom of the polarimeter assembly shown in Figure~\ref{fig:design}. A block diagram
of the DAQ system is shown in Figure~\ref{fig:daq-diagram}. The basic DAQ functionality is the same as that of the PoGOLite Pathfinder \cite{ppcalibration}.

PMT signals are amplified and continuously digitised at 100~MHz across six Flash Analog-to-Digital Converter (FADC) boards with 12-bit precision. 
A "fast-output" ("slow-output") is defined as the difference between the most recently sampled ADC value and that sampled 6 (37) clock cycles earlier, as dictated by the decay times of the plastic and BGO scintillators.
These quantities are used by the veto system to determine in which scintillator an interaction takes place. 
If the "fast-output" exceeds a user-defined trigger threshold a "level-0 trigger" (L0) is issued.

In comparison to the PoGOLite Pathfinder, a larger FPGA waveform buffer size is implemented which triples the maximum event registration rate. Consequently, PoGO+ has a trigger threshold which is $\sim50\%$ of that used for the PoGOLite Pathfinder. This results in an increase in effective detection area at higher energies making the instrument more sensitive to events comprising two low energy Compton scatterings.

A Digital Input/Output (DIO) board collects the L0 trigger signals from the FADC boards through a Logic Distribution Board (LDB). A "level-1 trigger" (L1) is returned in the absence of veto signals (e.g. from the anti-coincidence shield) initiating the storage of 10 pre-trigger and 40 post-trigger sample points (i.e. 500 ns window) for all read-out channels where the "fast-output" or "slow-output" exceeds a pre-set "hit" threshold (zero-suppression).

The PoGO+ data acquisition system has been improved in two additional ways. Firstly, on the FADC boards, a faster charge-sensitive amplifier input stage is used in conjunction with a faster FADC sampling frequency (100~MHz compared to 37.5~MHz). This potentially allows energy deposits from X-rays and fast neutrons to be distinguished in the plastic scintillator~\cite{Kamae}. Secondly, the power consumption of the DAQ system was decreased by $\sim30\%$, down to $\sim150$ W, in order to reduce demands on the fluid-based cooling system which also connects to the PMTs (which have integrated high-voltage power supplies). In particular, a single Xilinx Spartan-6 FPGA was used on each FADC board compared to the two Spartan-3 devices used previously.

In order to reconstruct the Crab pulsar light-curve, the absolute arrival time of X-rays is measured relative to Universal Time (UT) using a Pulse Per Second (PPS) timebase provided by the GPS unit within the attitude control system~\cite{ppcalibration}. Using an external GPS set-up, the precision of the PoGO+ timing system relative to UT is shown to be $\sim$100~ns.

\subsection{Online data reduction}
\label{sec:data_reduction}
In order to keep the measurement dead-time at an acceptable level an online veto system is implemented. Since 1-hit events do not carry any polarisation information but are abundant due to high background levels, they are discarded in flight (but are used for pre-flight calibration, see Section \ref{sec:energy_response}). As there is no direct communication between the FADCs, the LDB collects multiplicity information from the FADCs, identifies coincidences and generates a multiplicity signal ($>1$ hit) which is sent to the DIO. The DIO issues an L1 trigger only when the multiplicity signal is asserted, as shown in Figure~\ref{fig:daq-diagram}. For the PoGOLite Pathfinder, 1-hit events during the Crab observation constitute 21\% of all recorded waveforms after passing the online veto. Events with multiplicity $>2$ may be converted to 2-hit events offline, e.g. if the waveform baseline is not flat and is falsely identified as a hit.

Similarly to the PoGOLite Pathfinder, PoGO+ implements an upper discriminator (UD) veto for rejecting minimum ionising particles interacting in the scintillators. In order to achieve a more uniform response across SDCs with differing sensitivities, the SDC with the lowest UD threshold was identified, resulting in a UD threshold of 86~keV being applied to all SDCs. Following the same procedure, the SAS UD threshold was set to 1.75~MeV. 
For the PoGOLite Pathfinder, there were instances when the UD was not issued due to the saturation of the charge-sensitive amplifier on the FADC board, resulting in an elevated baseline and thus a "fast-" or "slow-output" which is too low to issue a UD signal. To prevent such occurrences, all 61 SDCs and 30 SAS units now issue a veto when the signal is higher than an absolute threshold value UDabs (set to the saturation level of 3100 ADC counts). Following large energy deposits by cosmic rays (which are discarded by the UD), the waveform baseline may undershoot and an additional lower discriminator threshold (LDabs) allows false signals caused by the recovering baseline to be immediately rejected. The DIO issues the L1 trigger only if no UD, UDabs, LDabs or waveform discriminator (WD) veto is issued and the multiplicity signal is asserted.

\begin{figure}[!th]
 \centering
   \includegraphics[width=0.5\textwidth]{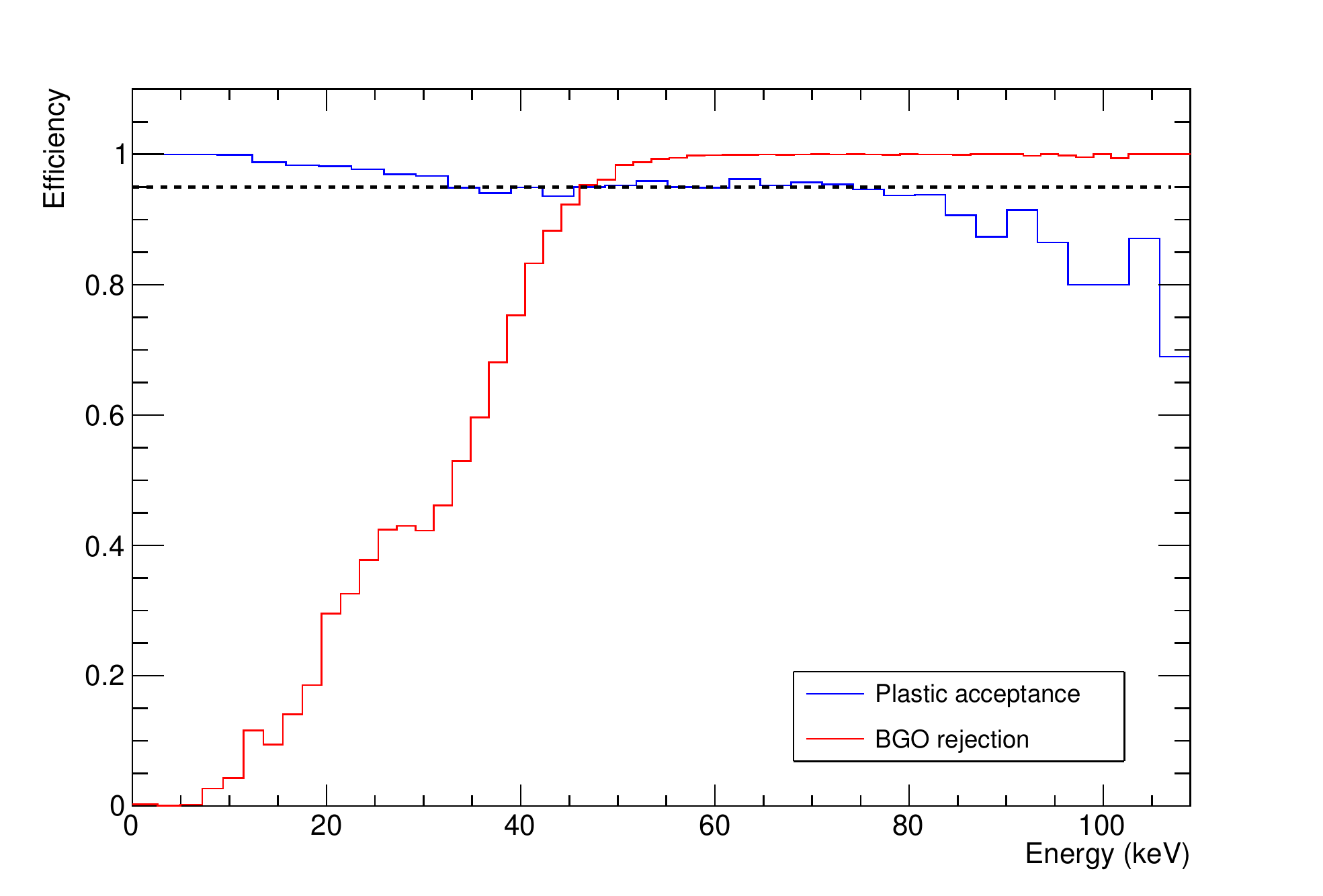} 
\caption{The acceptance of plastic (blue) and rejection of BGO (red) events as a function of incident energy by the online veto for an SDC with average sensitivity. The plastic acceptance decreases above 80 keV because the "fast-output" underestimates the waveform peak for high energies. The dashed line indicates 95\% efficiency.}
  \label{fig:Efficiency}
\end{figure}

The waveform selections used to define WD veto criteria are optimised in laboratory studies. A plastic scintillator is coupled to a BGO-dummy and a plastic-dummy is coupled to a BGO, where the dummy pieces are made from plexiglass and have the same shape as the corresponding scintillators. 
Since the dummy pieces are transparent to scintillation photons but do not themselves generate scintillation photons, these set-ups allow pure plastic and pure BGO waveforms to be studied, and the energy dependence of the waveform acceptance and rejection efficiency to be determined. 
Waveform selections are developed so that the plastic scintillator has an acceptance of at least 95\% up to 80 keV (a sufficiently high acceptance so that the system is not over-optimised which may be a problem if the flight and laboratory behaviour differ). 
The BGO rejection efficiency is optimised in a similar way.  Above 46 keV the BGO efficiency exceeds 95\%, as shown in Figure~\ref{fig:Efficiency} for a pure BGO waveform. The WD threshold is calculated as $8.45\times\mathrm{SPE}$, where SPE is the mean of the single photo-electron peak. A WD veto is issued if the value of the "slow-output" exceeds the sum of the "fast-output" and the WD threshold. Non-uniformities in the WD efficiency may arise from differences in SDC sensitivity, which has a relative spread of $\sim9\%$ around the central value of $0.87$ photo-electrons keV$^{-1}$.

To suppress electronic noise, promote a uniform response, and retain the events within the SPE peak, a hit threshold of $\mathrm{SPE}-2.5\sigma_{\mathrm{SPE}}$, but greater than 9 ADC counts, is set on all SDCs, where $\sigma_{\mathrm{SPE}}$ is the standard deviation of a Gaussian fit to the SPE spectrum for each SDC. 
A different approach is used for the SAS since a WD veto for SPE events would result in too much dead-time. Such events are likely due to dark current or low energy photons being stopped in the SAS and thus not reaching the SDCs. A hit threshold $\mathrm{SPE}+2\sigma_{\mathrm{SPE}}$ is set to suppress such vetoes. Compared to the PoGOLite Pathfinder flight, the PMT gain for SAS units is increased which results in improved rejection of high energy background photons which Compton scatter in the forward direction from the SAS to the SDCs.

The same trigger threshold energy is applied to all SDCs. Simulations show that the MDP improves as the trigger threshold is reduced down to 3~keV, making the instrument more sensitive to high energy photons undergoing two Compton scatterings in the polarimeter. However, the corresponding increase in dead-time means that this threshold needs to be fully optimised in flight.

\section{Simulation validation}
\label{sec:sim}

The value of $M_{100}$ for a celestial source is derived from simulations which describe the polarimeter response to the source energy spectrum once atmospheric attenuation is accounted for.
Particle interactions in the instrument and the atmosphere are simulated using Geant4 ~\cite{pogo+}. Then, the simulated interactions are convolved with the modelled detector response, which is based on laboratory measurements of individual detector components and is validated by comparing to data obtained from the entire polarimeter during calibration with radioactive sources.

\subsection{Energy response of individual components}
The test set-ups with [plastic scintillator and BGO-dummy] and [plastic-dummy and BGO scintillator], as described in Section~\ref{sec:data_reduction}, were recreated in simulations. Propagation of scintillation photons is not modelled, instead, an empirical model for the scintillator and PMT is created using several parameters: SPE, $\sigma_{\mathrm{SPE}}$, both measured in ADC counts, and the detector sensitivity measured in photo-electrons per keV of incident X-ray energy. The first two parameters are properties of the PMT and the third parameter describes the collective behaviour of a given PMT and scintillator. These parameters are measured by exposing each scintillator test set-up to a $^{241}$Am source with dominant emission at 59.5 keV. The simulation is developed from previous studies~\cite{KEKtest} where a prototype was tested at a synchrotron facility at energies of 30, 50 and 70 keV.

The $^{241}$Am source is encapsulated in a stainless steel container which absorbs most of the lower energy X-rays. A Germanium detector is used to measure a precise emission spectrum for use in the simulation. In order to achieve an acceptable agreement between simulation and measurement the incident energy used as an input parameter to the simulation for calculating the detector sensitivity is tuned to 56.5 keV since photons undergo Compton scattering in the steel container and in the passive material of the setup before reaching the plastic scintillator.

A correction is applied to scintillator energy deposits to compensate for non-linearities (quenching) in the production of scintillation light. 
The following empirical formulae are used for the plastic scintillator~\cite{nonlinMizuno} and the BGO~\cite{BGOnonlin}:
\begin{equation}
E_{\mathrm{plastic}}=E\times(0.216+0.500\arctan(0.393\times E^{0.590}))
\end{equation}
\begin{equation}
E_{\mathrm{BGO}}=E\times(0.598+0.248\arctan(0.160\times E^{0.638}))
\end{equation}
where $E$ is the incident photon energy in keV, and $E_{\mathrm{plastic}}$ ($E_{\mathrm{BGO}}$) are the deposited energies in the plastic (BGO) scintillators. 
A photon energy deposit of 56.5 keV is reduced by 11\% (12\%) in plastic (BGO) due to quenching, whereas a typical Compton scattering energy deposit of 6~keV is reduced by 36\% (29\%) in plastic (BGO).  
The attenuation of scintillation light along the length of the plastic scintillator uses the coefficient reported 
in~\cite{nonlinMizuno}, resulting in a 7.8\% reduction in detected energy for interactions occurring at the top of the scintillator. SAS units are treated in the same way as the bottom BGO but using a different attenuation length dependence, as described in~\cite{KissPhD}.

The simulated deposited energy in keV is converted into an equivalent number of photo-electrons (nPE). The value of nPE is smeared with a Poisson distribution to account for the statistics of emission at the photo-cathode.
To convert to ADC counts, including PMT amplification statistics, the nPE distribution is smeared using a Gaussian distribution with a mean of $\mathrm{nPE}\times \mathrm{SPE}$ and a standard deviation of $\sigma_{\mathrm{SPE}}\times\sqrt{\mathrm{nPE}}$.

\begin{figure}[!th]
 \centering
    \includegraphics[width=0.5\textwidth]{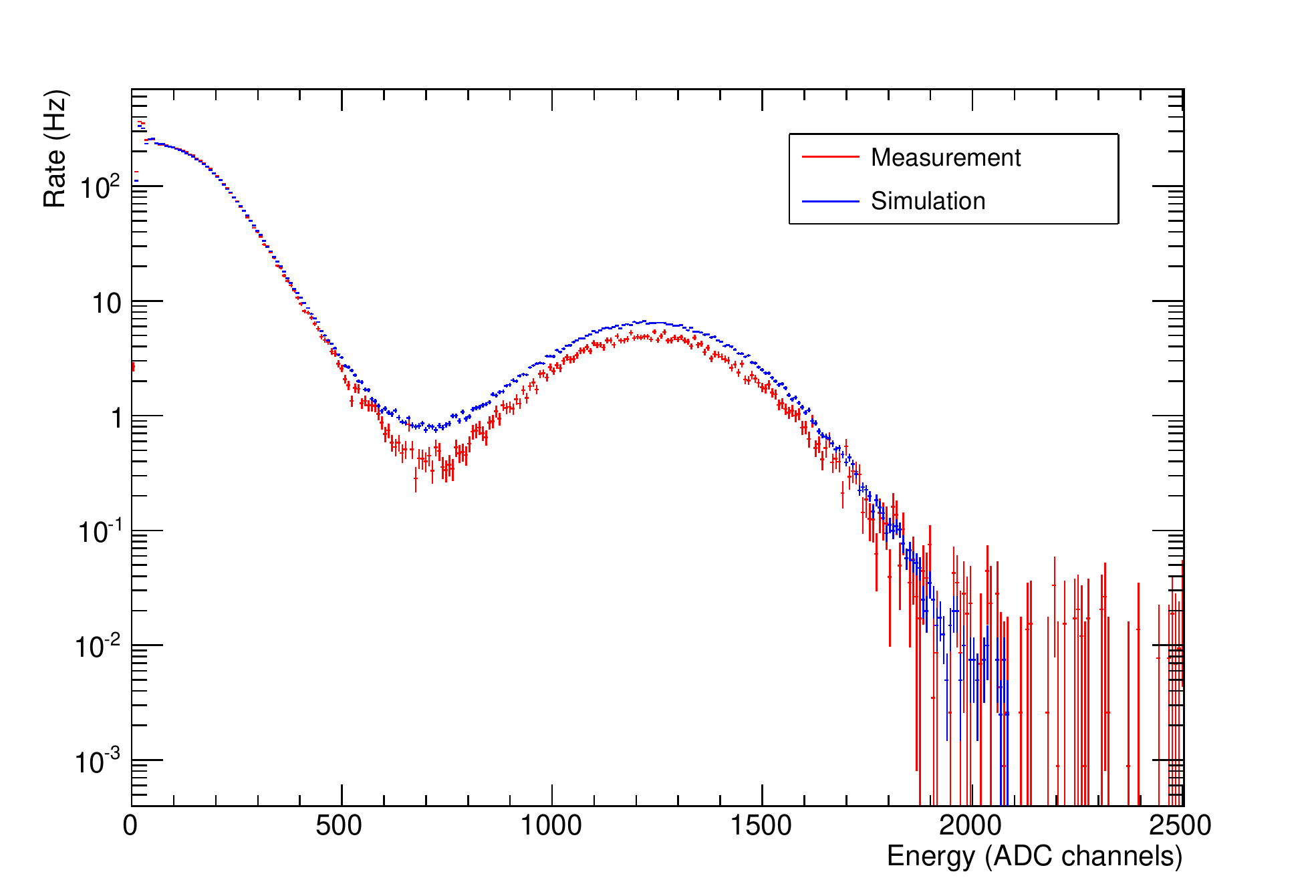}
    \includegraphics[width=0.5\textwidth]{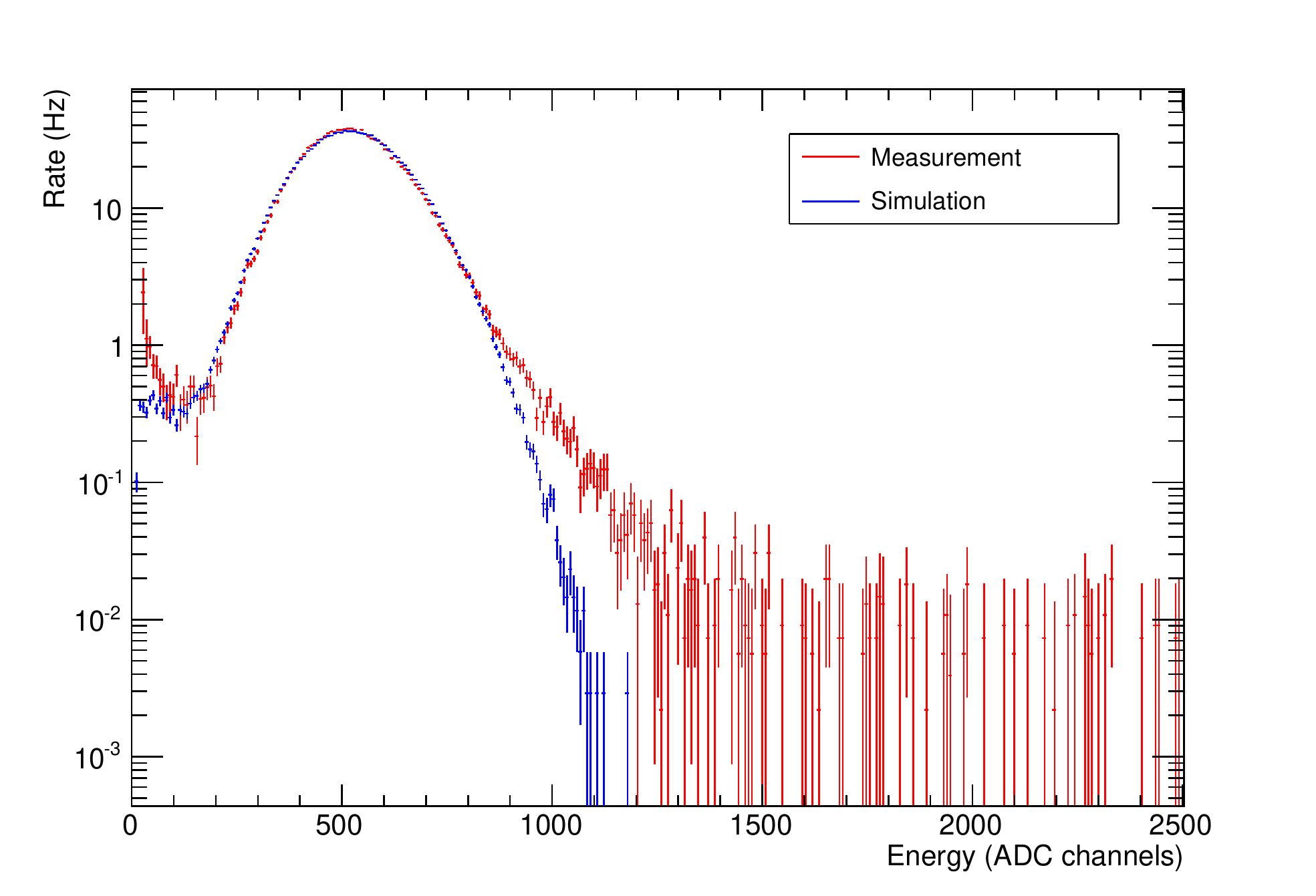}
   \includegraphics[width=0.5\textwidth]{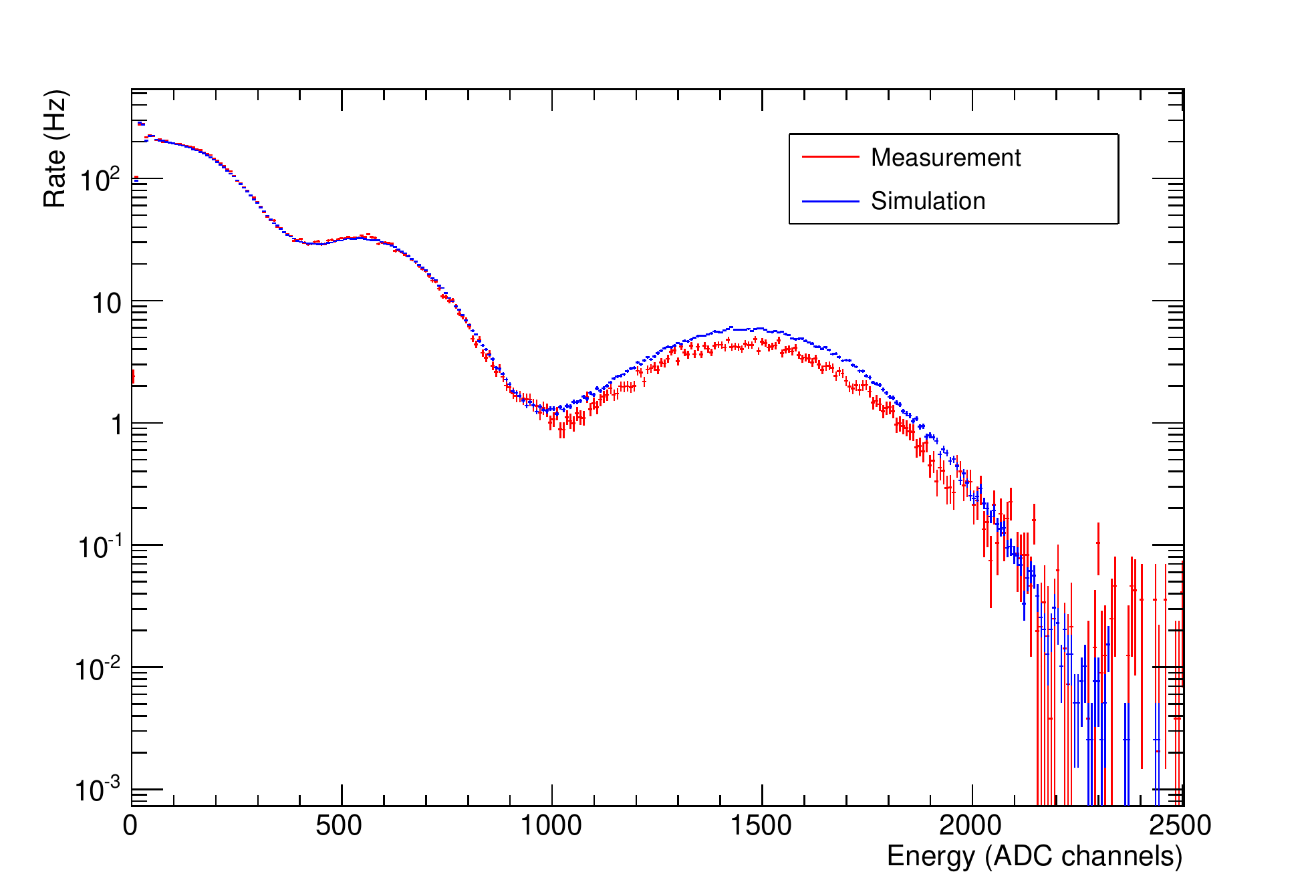} 
\caption{Data and simulation comparison. Top: Energy response for plastic scintillator and a plexiglass dummy with the shape of the BGO scintillator. Middle: Energy response for BGO and a plexiglass dummy with the shape of the plastic scintillator. Bottom: Energy response for a complete detector cell (SDC). At high energies the measurement has some residuals after background subtraction.}
  \label{fig:Eresponse}
\end{figure}

The resulting spectrum for the plastic scintillator is shown in the top panel of Figure~\ref{fig:Eresponse}. The simulated spectrum is normalised to the area of the background subtracted measured spectrum.
No further assumptions are made in the simulation.
The overall shape of the plastic scintillator spectrum is well reproduced by measurements. There are two reasons for the discrepancy in the ratio between low energy events (Compton scattering) and high energy events (photo-absorption). Firstly, the SPE distribution is assumed to be purely Gaussian, thereby neglecting a small (approximately) exponentially distributed low energy component which is produced when an electron does not go through all of the stages of acceleration, (e.g. photo-electrons missing the first dynode, photo-emission from the focusing electrodes and dynodes, etc.)~\cite{PMTresponse}. 
Secondly, the PMT is observed to exhibit after-pulsing, which is not treated in the simulation. To mitigate this in offline data analysis, the time since the previous "L0-trigger" from a given PMT is stored, allowing the removal of after-pulses by requiring a minimum time-of-arrival difference between consecutive events of 50 $\mu$s. Such a selection removes approximately 10\% of the 1-hit events. This approach is not fully efficient at the high source rates used for calibration measurements. After-pulsing has little impact on 2-hit events due to the short coincidence window of 110 ns.

The BGO spectrum is shown in the middle panel of Figure~\ref{fig:Eresponse}. 
As a consequence of after-pulsing, there are discrepancies in the low energy region.
The measured spectrum also shows a broad shoulder towards higher energies. 
The scintillation decay time of BGO (300~ns) corresponds to 30 waveform samples. Since the waveform starts at sample 10 for a total window of 50 samples, the peak of the BGO waveform may not be registered, thereby leading to an underestimation of the deposited energy. 

The bottom panel of Figure~\ref{fig:Eresponse} shows the energy spectrum for a complete SDC. Both the plastic and BGO photo-absorption peaks are clearly visible and there is good agreement between simulation and measurements. No additional tuning has been applied to the simulated spectrum (or any other spectrum presented here-after). The only tuned parameter is the incident energy used for the calculation of detector sensitivity, as described above.

\subsection{Polarimeter energy response}
\label{sec:energy_response}

\begin{figure}[!th]
 \centering
    \includegraphics[width=0.5\textwidth]{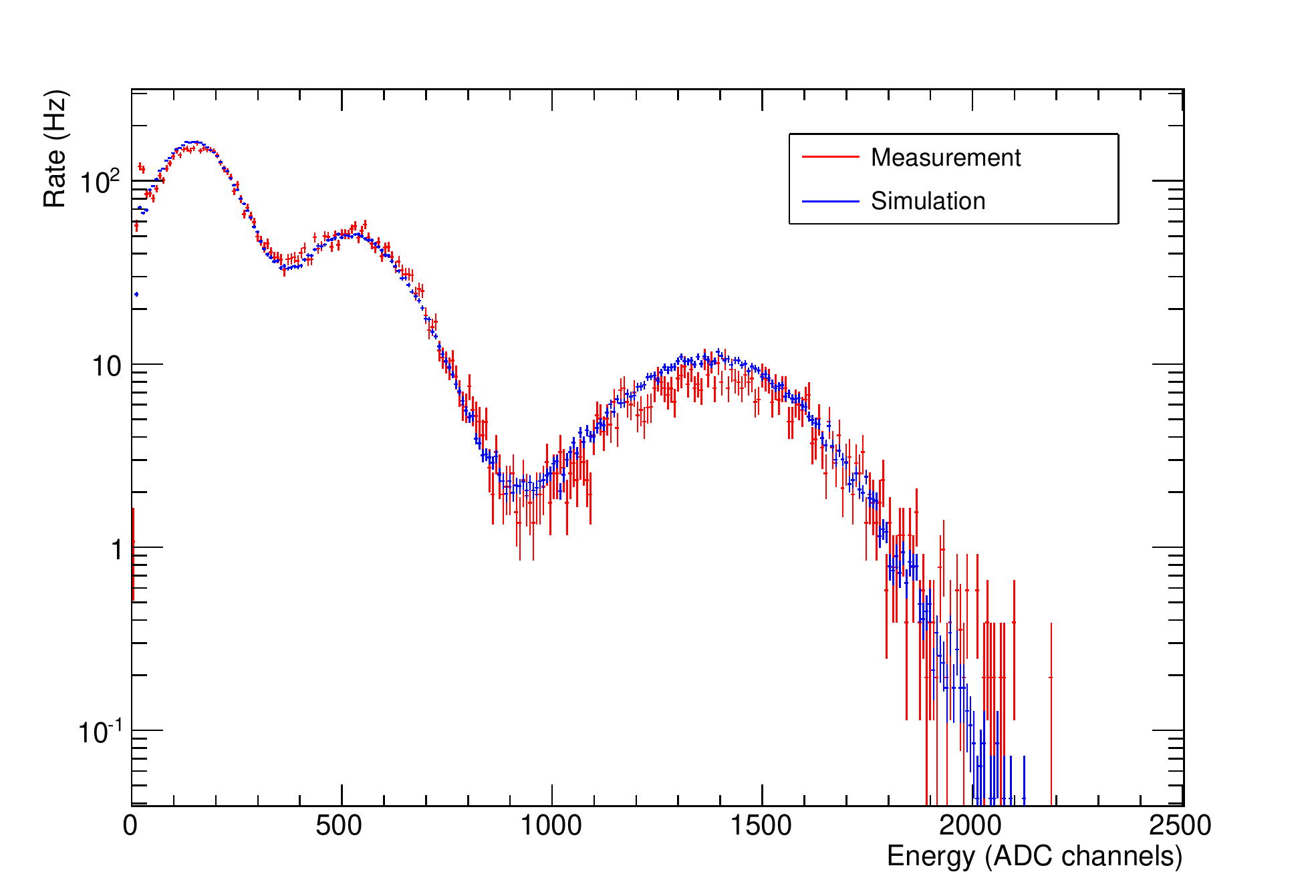}
\caption{Data and simulation comparison. Spectrum for 1-hit events in the entire polarimeter when the central unit is irradiated from the front. The three broad peaks, from right to left, are photo-absorption in the plastic scintillator, photo-absorption in the BGO and Compton scattering in the plastic scintillator. The narrow peak on the left is the SPE peak.}
  \label{fig:1hit}
\end{figure}

In order to study the 1- and 2-hit spectrum, hit and trigger thresholds are set to 10 ADC counts and the online veto is disabled.
The central SDC of the polarimeter is irradiated with the $^{241}$Am source. 
Figure~\ref{fig:1hit} shows the resulting 1-hit spectrum (events with higher multiplicity are discarded). 
The broad peak immediately to the right of the SPE peak arises from Compton scattering in the plastic scintillator. 
A peak is formed rather than a continuum because only a very limited range of scattering angles is allowed for the 1-hit criteria. 
The agreement is good in all parts of the spectrum except the SPE peak, where the discrepancy is most likely due to after-pulsing.

\begin{figure}[!th]
 \centering
    \includegraphics[width=0.5\textwidth]{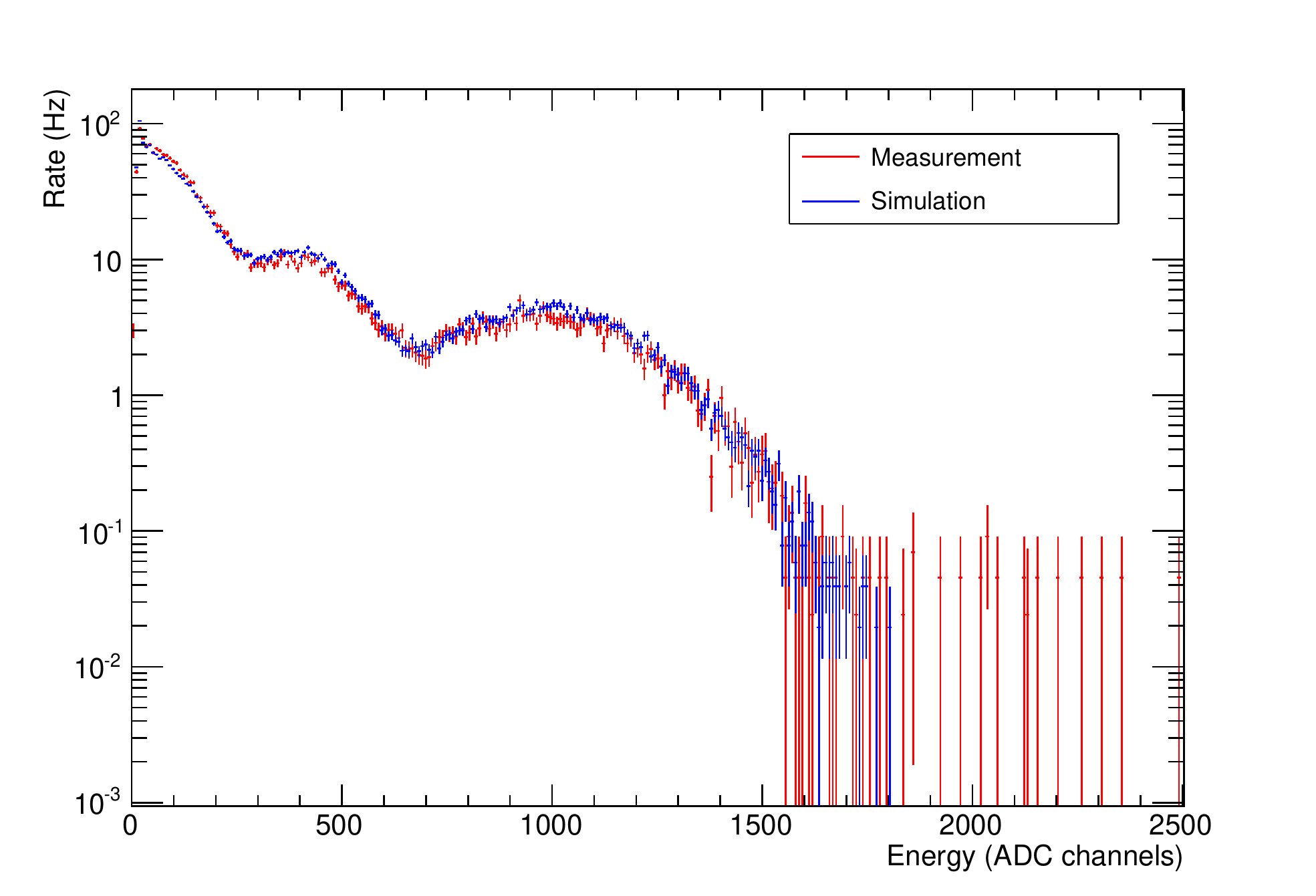}
    \includegraphics[width=0.5\textwidth]{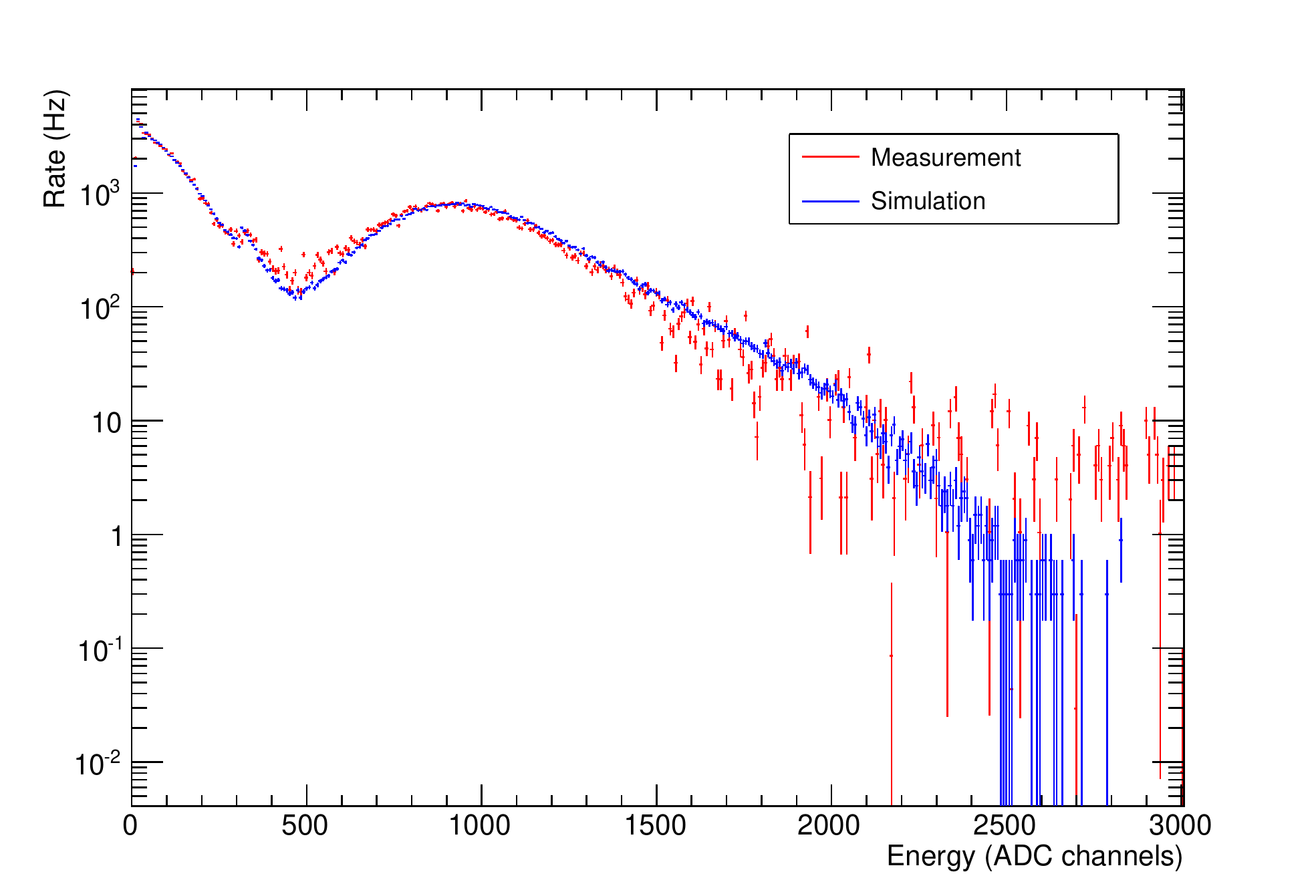}
\caption{Data and simulation comparison. The central unit of the instrument is irradiated from the front. The energies of the individual hits in a 2-hit event are recorded. Top: 2-hit spectrum for the central unit and the first ring of the instrument. All vetoes are off and the trigger and hit thresholds are set at 10 ADC counts. Bottom: 2-hit spectrum for all rings. All vetoes are on and the thresholds are set as foreseen for flight.}
  \label{fig:2hit}
\end{figure}

The 2-hit spectrum (which codes polarisation information) is shown in Figure~\ref{fig:2hit}. The top panel shows the energy of the individual hits in the central SDC and the first ring of SDCs. There is good agreement between the distributions at all energies.

\subsection{Thresholds and online veto}
\label{sec:sim_veto}

The simulation does not model the PMT waveform shape.
Instead, the online veto is simulated by assigning a "fast-" and "slow-output" to each SDC. 
Interactions in the plastic scintillator generates comparable "fast-" and "slow-output" values, whereas BGO interactions result in a "fast-output" which is $3$ times lower than the "slow-output". 
The resulting "fast-" and "slow-output" values are compared to the online veto condition.
Thresholds described in Section~\ref{sec:data_reduction} are also applied. The lower panel of Figure~\ref{fig:2hit} shows the resulting 2-hit spectrum for the entire polarimeter.

\subsection{Energy range for polarisation measurements}

\begin{figure}[!th]
 \centering
    \includegraphics[width=0.5\textwidth]{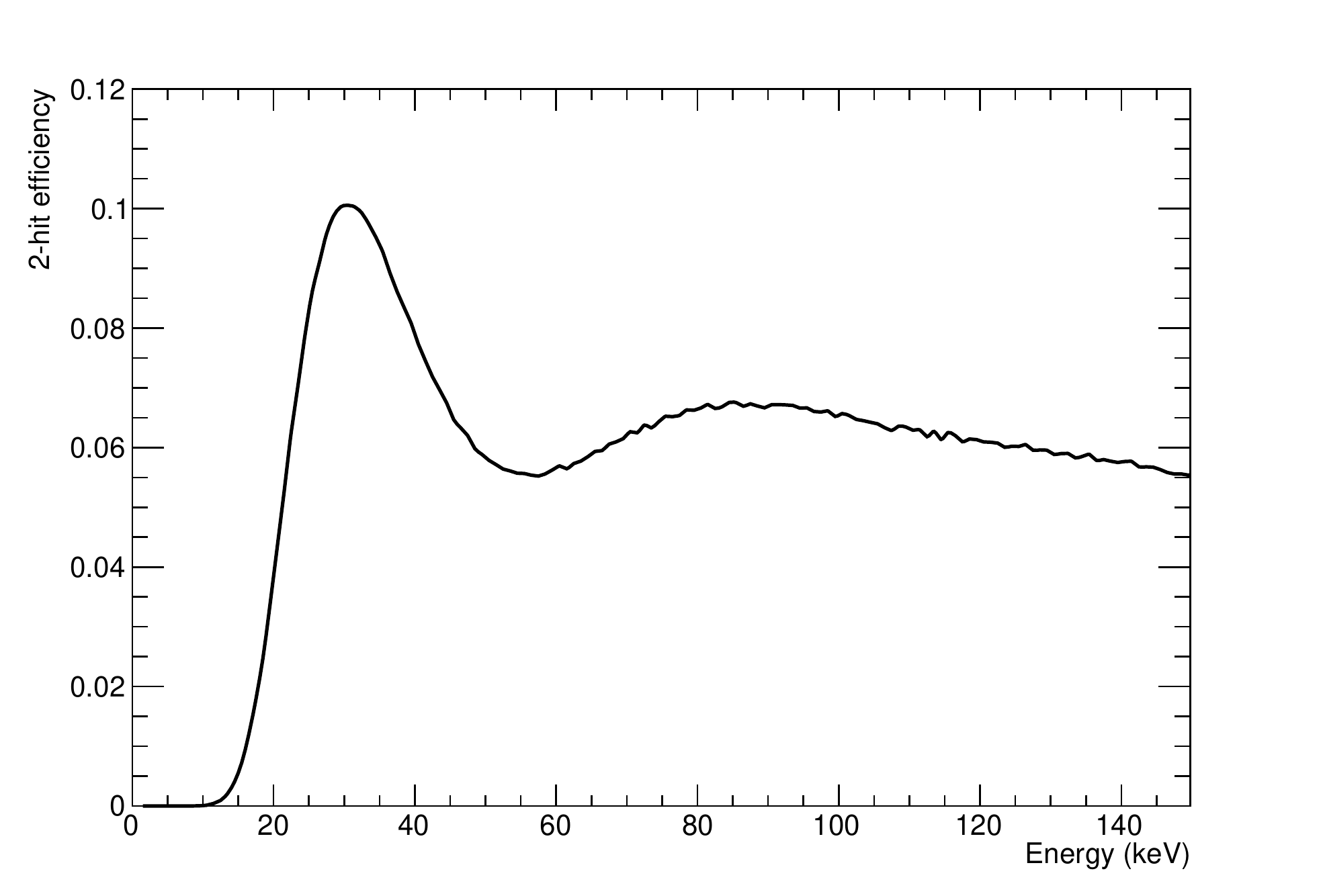}
    \includegraphics[width=0.5\textwidth]{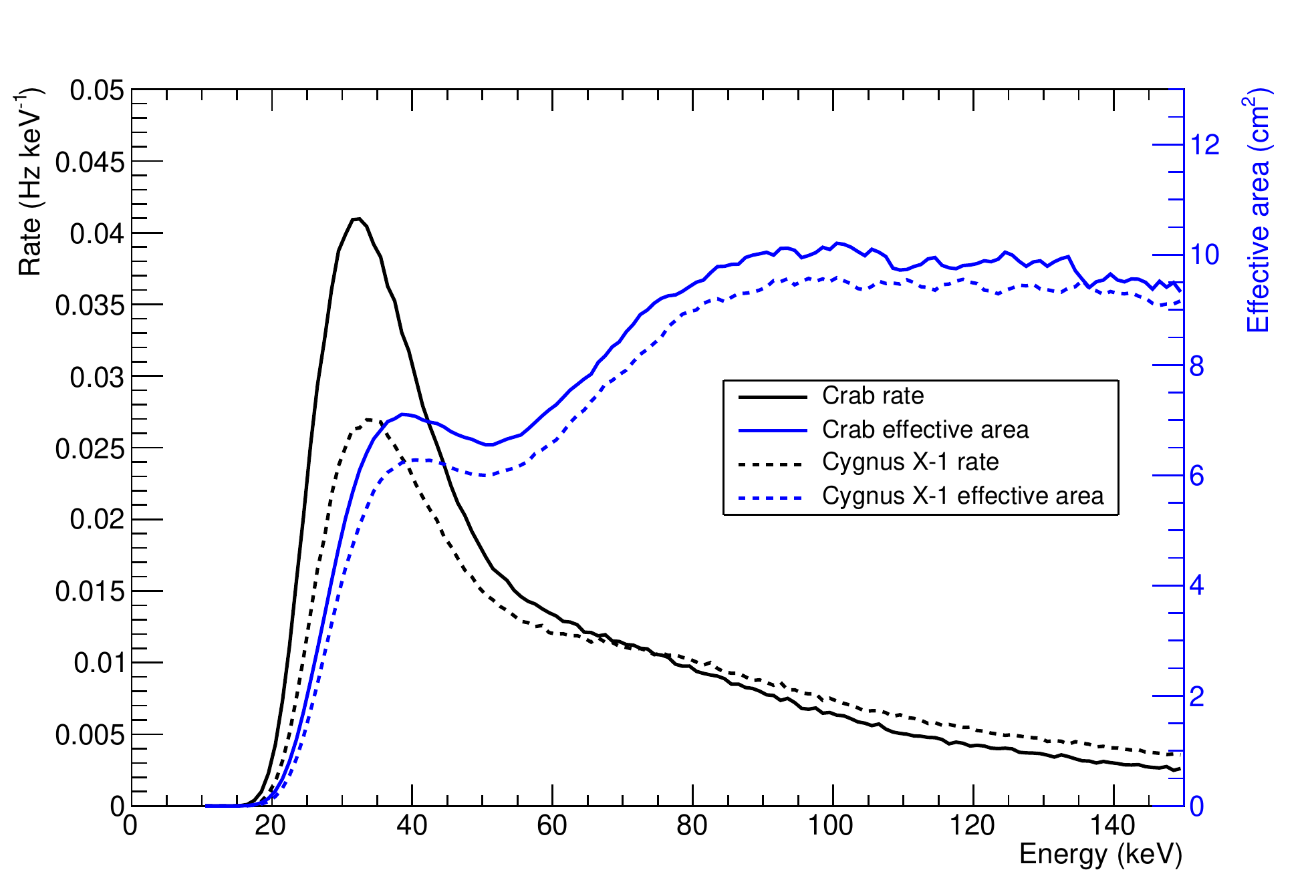}
\caption{Top: the simulated 2-hit efficiency as a function of incident photon energy for the entire polarimeter assuming no atmospheric attenuation. Bottom: The effective area and simulated 2-hit event rate as a function of photon energy at source for the Crab and Cygnus X-1. The simulation accounts for atmospheric absorption and source spectral shape. Limited simulation statistics lead to fluctuations above 80 keV.}
  \label{fig:rate_plot}
\end{figure}

The simulated instrument response is used to predict the in-flight performance for the Crab and Cygnus X-1. The absolute 2-hit detection efficiency for the entire polarimeter as a function of the incident photon energy is shown in the top panel of Figure~\ref{fig:rate_plot}. The first peak at 30 keV corresponds to polarisation events where incoming photons are first Compton scattered in one detector and subsequently photo-absorbed in another, while the second broad peak (around 85 keV) is caused by photons that Compton scatter twice before being absorbed in passive materials.

The bottom panel of Figure~\ref{fig:rate_plot} shows the effective area as a function of energy where atmospheric attenuation is taken into account assuming the average column density of the PoGOLite Pathfinder flight, $5.4$ ($6.7$) g cm$^{-2}$ for the Crab (Cygnus X-1). The rate at which the polarimeter detects source photons is also shown using the source spectra $9.7E^{-2.1}$~\cite{CrabSpec} and $2.0E^{-1.7}$~\cite{CygSpec} photons cm$^{-2}$ s$^{-1}$ keV$^{-1}$ for the Crab and Cygnus X-1, respectively. For the Crab (Cygnus X-1), the median energy is 51 (67) keV, which supports the suitability of $^{241}$Am as a calibration source.

Polarisation measurements are made where the rate exceeds 5\% of the maximum for the Crab. Simulations show that this corresponds to an energy range of 20--150 keV. The lower energy limit is due to atmospheric attenuation, while the high energy limit is dictated by expected background conditions.

\section{Polarimeter response}
\label{sec:response}

The response of the polarimeter to both unpolarised and polarised radiation has been determined prior to flight. 
Characterising the response to unpolarised radiation is especially important due to the positive-definite nature of polarisation measurements, where variations in detector response may result in a measured net polarisation even for an unpolarised photon flux.

The instrument calibration strategy is similar to that described in \cite{ppcalibration} for the PoGOLite Pathfinder mission, but with one major improvement -- source and background measurements are interspersed.
This is important because during the lengthy measurements, the background rate can change significantly ($\sim5\%$ level), e.g. due to day and night temperature variations in the electronics. 
Asymmetries can also be introduced during laboratory tests, e.g. due to the presence of passive structural materials, which can result in a small but measurable background polarisation fraction ($\sim0.4\%$ for the test set-up described below). Therefore robust background subtraction is a prerequisite for an accurate calibration measurement.

\subsection{Analysis}
 
Only 2-hit events are used for determining the distribution of scattering angles. The modulation curve is shown in Figure~\ref{fig:modcurve} for both polarised and unpolarised beams from a $^{241}$Am source.
The modulation curve is primarily used to study possible deviations from the expected harmonic behaviour.
Each scattering angle, $\phi_i$, is calculated by subtracting the polarimeter roll angle from the angle between the SDCs where the 2-hit interactions take place. 
Some roll angles are over-represented since measurements start before the roll motor is powered. 
A weight $w_i$ is assigned to each event, to produce a flat histogram of roll angles.  
These weights are applied to the histogram of scattering angles.
The statistical (Gaussian) uncertainty for each bin is calculated from the unweighted histogram. 
The same procedure is applied to the background measurement and the resulting scattering angle histogram is subtracted from the source histogram bin-by-bin and the errors are added in quadrature. 

In order to avoid binning effects, polarisation parameters are determined using Stokes parameters, defined as:
\begin{equation}
\begin{split}
Q=\sum_{i=1}^N w_i\cos{2\psi_i}; \phantom{bla} U=\sum_{i=1}^N w_i\sin{2\psi_i}; \\  I=\sum_{i=1}^N w_i;\phantom{bla}  W^2=\sum_{i=1}^N w_i^2
\end{split}
\end{equation}
where $N$ is the total number of measured photons. Based on derivations~\cite{StokesKislat} for weighted Stokes parameters, the background-subtracted modulation factor $M$, polarisation angle $\psi$ and their respective uncertainties $\sigma_M$ and $\sigma_\psi$ are given by

\begin{equation}
M=\frac{2}{I_{\mathrm{src}}-I_{\mathrm{bg}}}\sqrt{(Q_{\mathrm{src}}-Q_{\mathrm{bg}})^2 + (U_{\mathrm{src}}-U_{\mathrm{bg}})^2}
\end{equation}

\begin{equation}
\sigma_M=2\sqrt{\bigg(\frac{I_{\mathrm{src}}+I_{\mathrm{bg}}}{2(I_{\mathrm{src}}-I_{\mathrm{bg}})}-\frac{M^2}{4}\bigg)\times\frac{W^2_{\mathrm{src}}+W^2_{\mathrm{bg}}}{(I_{\mathrm{src}}+I_{\mathrm{bg}})(I_{\mathrm{src}}-I_{\mathrm{bg}})}}
\end{equation}

\begin{equation}
\psi=\frac{1}{2}\arctan{\frac{U_{\mathrm{src}}-U_{\mathrm{bg}}}{Q_{\mathrm{src}}-Q_{\mathrm{bg}}}}
\end{equation}

\begin{equation}
\sigma_\psi=\frac{1}{M\sqrt{2(I_{\mathrm{src}}-I_{\mathrm{bg}})}}\times\sqrt{\frac{W^2_{\mathrm{src}}+W^2_{\mathrm{bg}}}{I_{\mathrm{src}}-I_{\mathrm{bg}}}}
\end{equation}
where all of the Stokes parameters for the background are normalised for the live-time. The modulation curve phase $\phi_0$ and the polarisation angle are related by $\psi=\phi_0-90^\circ$.

\subsection{Response to unpolarised beams}

To determine the response to an unpolarised beam the $^{241}$Am source was placed in a collimating holder on top of the polarimeter, illuminating the central unit of the detector array. 
An electromagnetic shutter allowed automated alternating source and background observations. 
During measurements, the detector array was rolled back-and-forth as foreseen for flight. 
Each source or background measurement includes two such revolutions, so that the array rotates four times during one measurement cycle (i.e. 24 min cycles).
The modulation curve of the central unit (after background subtraction) is shown in Figure~\ref{fig:modcurve}. 
Data was acquired with all vetoes on. 
A Stokes analysis yields a modulation factor of $(0.10\pm0.12)\%$, consistent with the expectation for an unpolarised source. 
Since $\chi^2/\mathrm{d.o.f}=133/177$, the fit to Equation~\ref{eqn:modcurve} appears over-determined. Systematic studies of data selections and error determinations did not yield any evidence for error over-estimation. 

\subsection{Response to polarised beams}

\begin{figure}[!th]
\hspace*{-0.3cm}
 \centering
    \includegraphics[width=0.5\textwidth]{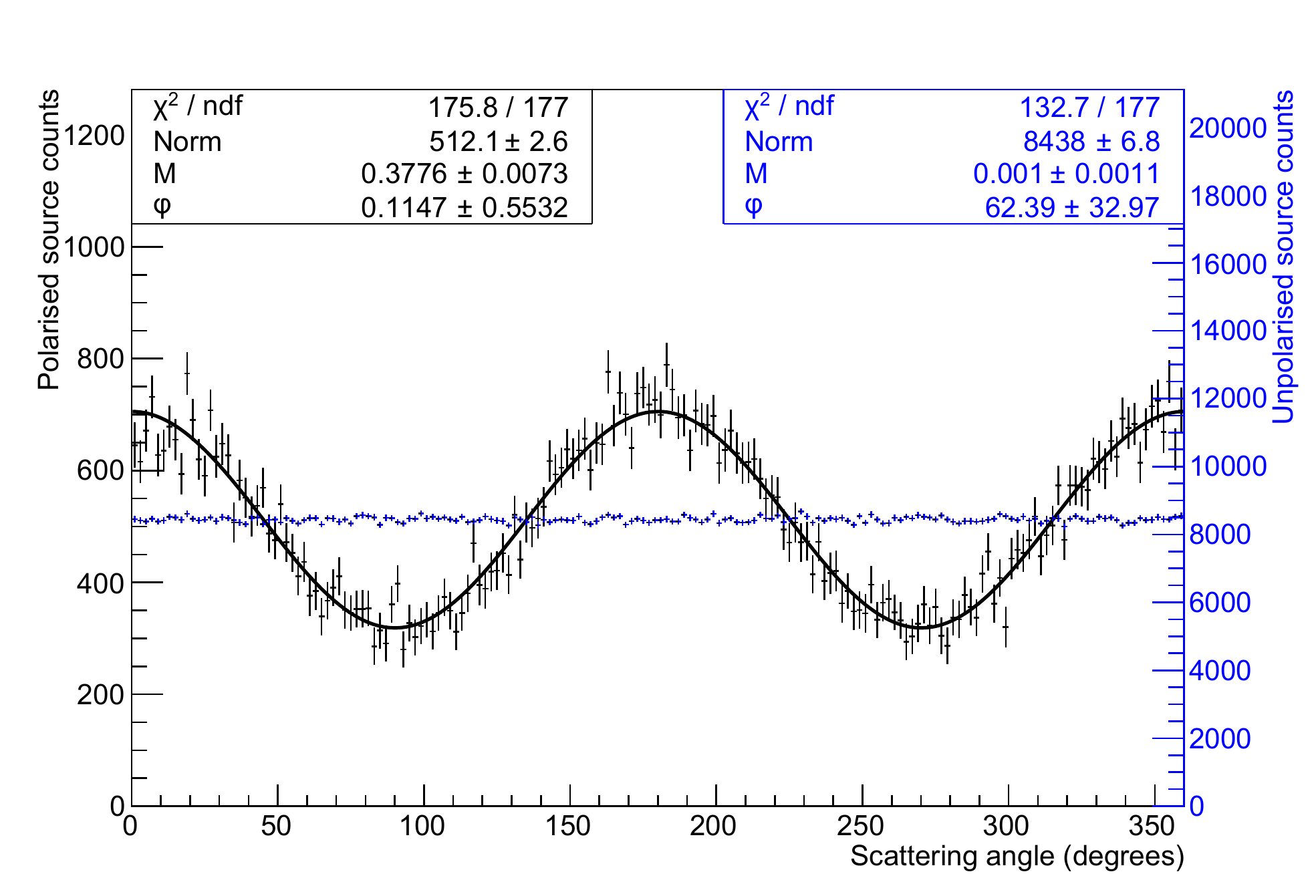}
\caption{The modulation curve reconstructed for a polarised $^{241}$Am source (black) and unpolarised $^{241}$Am source (blue).}
  \label{fig:modcurve}
\end{figure}

To create a polarised source, a piece of polyethylene was placed next to the aperture of the $^{241}$Am source. 
Photons scattering through $90^\circ$ ($\sim100\%$ polarised), were directed towards the central SDC in the polarimeter.
The same data acquisition procedure was followed as for the unpolarised case and the resulting modulation curve is shown in 
Figure~\ref{fig:modcurve}.

The modulation curve is well described by Equation~\ref{eqn:modcurve} ($\chi^2/\mathrm{d.o.f}=176/177$). A Stokes analysis yields a modulation factor $M_{\textrm{meas}}=(37.83\pm0.73)\%$. The resulting polarisation angle ($0.14\pm0.56$)$^\circ$ is consistent with $0^\circ$, as expected from the measurement set-up. 
Simulations of the instrument response to the source and scattering piece, performed using the settings described in Section~\ref{sec:sim_veto}, yield $M_{\textrm{sim}}=(40.04\pm0.25)\%$. 
The measured modulation factor is lower than the simulated value, most likely because the simulation does not capture all details of the polarimeter construction and measurement set-up. 
The scaling factor between measurement and simulation is $0.945\pm0.019$. Using this factor, the simulated modulation factor for the Crab $M_{\textrm{Crab,sim}}=(44.19\pm0.12)\%$ can be converted to an expected value of $M_{\textrm{Crab}}=(41.75\pm0.85)\%$. This is a significant improvement over the PoGOLite Pathfinder where $M_{\mathrm{Crab}}=(21.4\pm1.5)\%$~\cite{ppresults}.

\subsection{Predicted performance}

\begin{figure}[!th]
 \centering
    \includegraphics[width=0.5\textwidth]{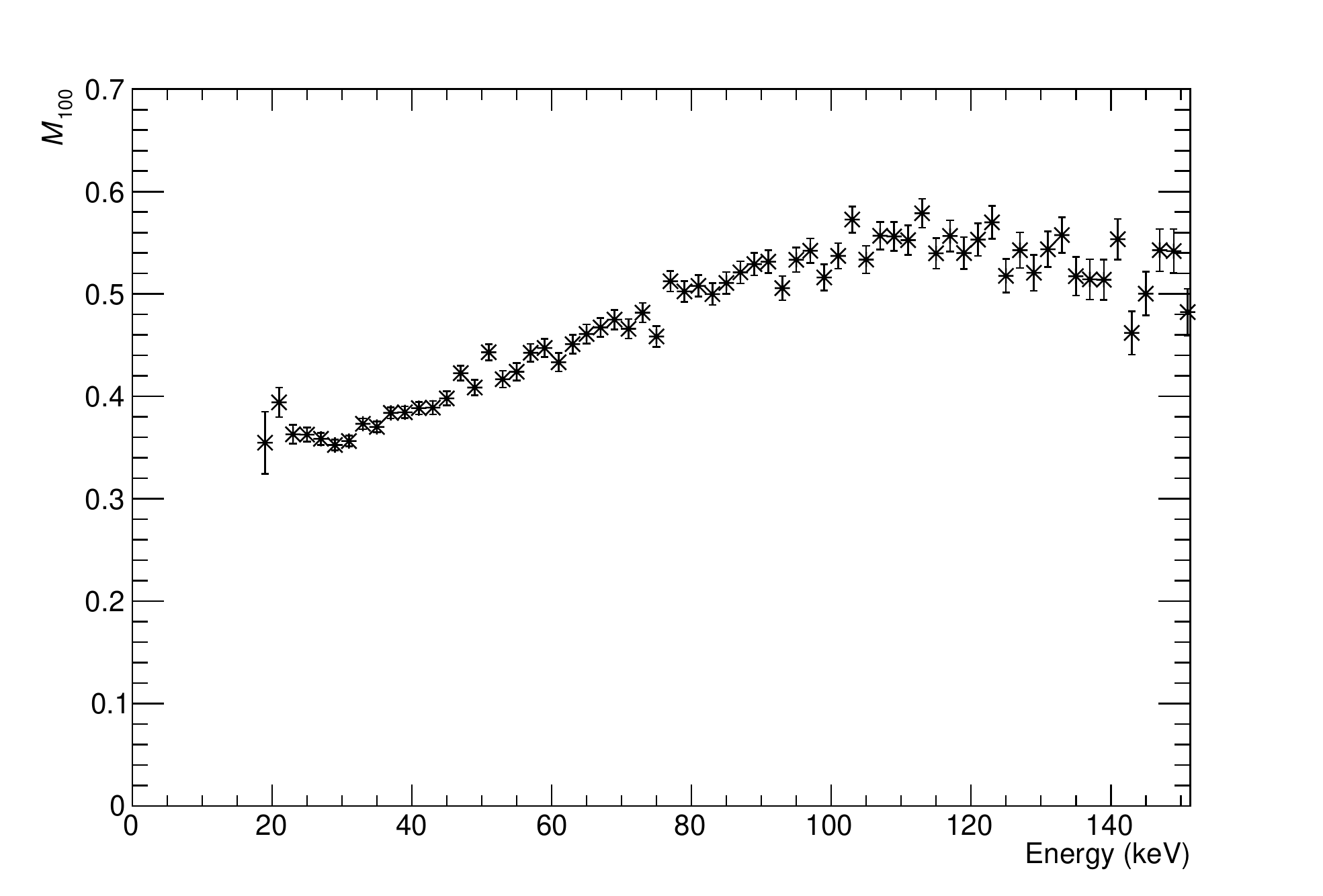}
\caption{The simulated M$_{100}$ as a function of energy. Data points below $\sim20$~keV are not simulated due to atmospheric attenuation.}
 \label{fig:M100}
\end{figure}

Figure~\ref{fig:M100} shows the simulated energy dependence of $M_{100}$. 
The value of $M_{100}$ increases up to 120~keV reflecting that photons travel farther on average between two interaction sites, thereby constraining the scattering angle.
Above 120 keV $M_{100}$ decreases because the scattering angle is less correlated to the polarisation angle, as described by the Klein-Nishina formula.

Combining the results from Figures~\ref{fig:rate_plot} and~\ref{fig:M100} with background simulations yields an MDP of $7.3\%$ for the Crab for a 7 day flight given a 50\%/50\% on-source/off-source measurement strategy (optimal for low signal-to-background ratio instruments~\cite{StokesKislat}), which is better than the $9.4\%$ predicted in the original PoGO+ design study~\cite{pogo+}. Such an observation strategy allows an estimation of the background rate and enables the subtraction of possible background modulation. The increase in performance is due to further improvements to the instrument design, e.g. lowering the trigger threshold to 10 keV, the addition of extra polyethylene shielding to balance the instrument within the attitude control system, and increasing the gain of SAS PMTs, as described in Section~\ref{sec:data_reduction}.

Background measurements may show a statistically significant modulation. For the 50\%/50\% on-source/off-source measurement strategy, the resulting background subtraction increases the MDP by a factor $\sim\sqrt{2}$ to $\sim10.3\%$. Previously, PoGOLite Pathfinder measured a polarisation fraction $\Pi=(18.4^{+9.8}_{-10.6})\%$. For the same polarisation fraction, PoGO+ will be able to make a measurement with $\sim5.4\sigma$ precision where
\begin{equation}
\sigma=\frac{2}{M_{100}}\sqrt{\frac{1}{I_{\mathrm{src}}}\bigg(\frac{I_\mathrm{src}+I_\mathrm{bg}}{2I_\mathrm{src}}-\frac{M_{100}^2\Pi^2}{4}\bigg)}
\end{equation}
Similar performance is expected for Cygnus X-1 (hard spectral state).

The above approximation does not consider uncertainties on the modulation factor or the signal-to-background ratio, nor does it follow the necessary Bayesian approach in the low statistics regime (see discussion in~\cite{Quinn,Maier}), but illustrates the expected performance.


\section{Summary and outlook}

PoGO+ is a Compton polarimeter for celestial sources where the design builds on experience gained during the 2013 balloon flight of the PoGOLite Pathfinder. A previous publication~\cite{pogo+} has detailed Geant4 simulation studies which led to a new instrument design for PoGO+ e.g.: the use of passive collimators; shortening the length of the plastic scintillators, and improving their light yield; eliminating optical cross-talk between detector units; and improving the hermeticity of the neutron shield. In addition, the data acquisition system has been significantly improved through an increase in PMT waveform sampling rate, from 37.5~MHz to 100~MHz, and waveform buffer size; and the introduction of a more comprehensive veto system. These measures decrease dead-time and improve the background rejection efficiency. Finally, the polarimeter calibration methodology has also been improved, resulting in a more uniform energy response and rejection/acceptance efficiency across all SDCs and SAS units. Laboratory tests have demonstrated that the polarimetric performance of the PoGO+ flight instrument exceeds expectations from previously published simulation studies. A new Geant4 simulation model of the instrument has been validated by comparing the energy spectra and polarimetric performance with laboratory measurements. The validated simulation is then used to calculate the modulation factor $M_{\mathrm{Crab}}=(41.75\pm0.85)\%$ and predict an expected performance of $\mathrm{MDP}=7.3\%$ for a 7 day flight which, after background subtraction, results in a statistical significance of better than $5\sigma$ ($\mathrm{MDP}=10.3\%$) assuming the same Crab polarisation fraction as measured by the PoGOLite Pathfinder~\cite{ppresults}. A similar performance, in terms of MDP, is expected for Cygnus X-1 in the hard spectral state.

Phase selections on the Crab light-curve provide a means of separating the nebula from the pulsar, and results in a MDP of $\sim 19\%$ (after background subtraction) for the nebula. A comparable polarisation fraction has been previously reported for the nebula at 5.2 keV~\cite{OSO8}. PoGO+ will therefore make a measurement constrained to at least 3$\sigma$-level, although it is noted that the polarisation fraction is expected to increase with energy.

\section{Acknowledgements}

PoGO+ is developed using funding provided by The Swedish National Space
Board, The Knut and Alice Wallenberg Foundation, The Swedish Research
Council and JSPS KAKENHI (grant number JP25302003).
Contributions from M.Sc.\,Thesis students at KTH (P.~Ekfeldt,
L.~Eliasson, H.~Wennl\"{o}f and R.~\"{O}stlund) are gratefully acknowledged.
R.~Helg and K.~H\"{o}rnfeldt at AlbaNova University Centre are thanked for
their help in constructing the polarimeter.
M.~Kole and E.~Moretti made important contributions to many aspects of the PoGOLite Pathfinder mission from which PoGO+ is derived. M.~Arimoto, Y.~Fukazawa, T.~Kamae, J.~Kataoka, N.~Kawai, T.~Mizuno, H.~Tajima and T.~Takahashi are thanked for their early work on the PoGOLite concept and for providing PMTs and some parts of the data acquisition.
H.~Tajima also assisted with the redesign of the Flash ADC boards. DST
Control developed the PoGO+ attitude control system which was used during
some of the measurements reported here. SSC personnel (and C.~Lockowandt in
particular) are thanked for practical assistance during tests conducted at
Esrange Space Centre.

\end{document}